\documentclass[referee]{aa}

\usepackage{graphicx}
\usepackage{natbib}
\usepackage{txfonts}

\begin{document}

\title {The chemical evolution of Manganese in different stellar systems}

\author {G. Cescutti\inst{1}
\thanks {email to: cescutti@ts.astro.it}
\and  F. Matteucci\inst{1, 2}
\and G.A. Lanfranchi\inst{3}
\and A. McWilliam\inst{4}}
\institute{Dipartimento di Astronomia, Universit\'a di Trieste, via G.B. Tiepolo 11, I-34131  
\and  I.N.A.F. Osservatorio Astronomico di Trieste, via G.B. Tiepolo 11, I-34131
\and  N\'ucleo de Astrof\'\i sica Te\'orica, Universidade
Cruzeiro do Sul, R. Galv\~ao Bueno 868, Liberdade, 01506-000, S\~ao Paulo, SP, Brazil
\and Observatories of the Carnegie Institution of Washington, 813 Santa Barbara St., Pasadena, CA, USA
}
\date{Received xxxx / Accepted xxxx}

\abstract{}{To model the chemical evolution of manganese relative to iron in three 
different stellar systems:
the solar neighbourhood, the Galactic bulge and the Sagittarius dwarf spheroidal galaxy,
and compare our results with the recent and homogeneous observational data.}
{We adopt three chemical evolution models well able to reproduce the main properties
of the solar vicinity, the galactic Bulge and the Sagittarius dwarf spheroidal.
Then, we compare different  stellar yields in order to identify the best set
to match the observational data in these systems.}
{We compute the evolution of manganese in the three systems and we find that in order to reproduce 
simultaneously the [Mn/Fe]
versus [Fe/H] in the Galactic bulge, the solar neighbourhood and Sagittarius,  
the type Ia SN Mn yield must be metallicity-dependent.}
{We conclude that the different histories of star formation in the three systems are not enough to reproduce the 
different behaviour of the [Mn/Fe] ratio, unlike the situation for [$\alpha$/Fe]; rather, it is necessary to
invoke metallicity-dependent type Ia SN Mn yields, as originally suggested by McWilliam, Rich \& Smecker-Hane in 2003.}


\titlerunning{The chemical evolution of Manganese in different stellar systems}

\maketitle

\authorrunning{Cescutti et al.}

\section{Introduction}
McWilliam, Rich \& Smecker-Hane (2003) considered the abundance trend of Mn in three contrasting stellar populations: a sample of Galactic 
bulge K giants, stars in the solar neighbourhood and stars in the dwarf spheroidal galaxy Sagittarius (Sgr dSph).
They found that the trend of [Mn/Fe] in the Galactic bulge follows more or less the relation of the solar vicinity,
but most stars of Sgr dSph show [Mn/Fe] deficient by $\sim 0.2$ dex. They concluded that to explain this result one has
to assume that the yields of Mn both from type II and type Ia SNe should depend on the initial stellar metallicity. 
This conclusion was at variance with a suggestion by Gratton (1989) that Mn is overproduced in type Ia SNe, relative to 
type II SNe.
Gratton (1989) observed that [Mn/Fe] versus [Fe/H] in solar vicinity stars show the opposite behaviour of the [$\alpha$/Fe] 
ratios: in fact, [Mn/Fe] $\sim$$-$0.4 dex and constant for stars with [Fe/H]$<-1.0$ dex and increases up to the solar value for 
stars with [Fe/H]$>$ $-$1.0 dex.
As it is well known, first Tinsley (1979) and then Greggio \& Renzini (1983) and Matteucci \& Greggio (1986) explained
the trend of the $\alpha$-elements (e.g. O, Mg, Si, Ca, Ti), relative to Fe, as due to the different roles played by 
type Ia and II SNe in galactic 
chemical enrichment. In particular,  the higher than solar [$\alpha$/Fe] ratio in Halo stars is interpreted as 
due to the pollution of type II SNe, which explode on very short timescales (of the order of Myr to tenths of Myr)
and produce 
roughly constant [$\alpha$/Fe] ratios. Then the [$\alpha$/Fe] ratio decreases to the solar value in disk stars because of the
bulk of Fe produced by type Ia SNe explode on timescales ranging from 30--40 Myr to 10 Gyr. 
This interpretation is known as ``time-delay model'' and represents the best interpretation of 
alpha-element abundance ratios in galaxies so far.
Type II SNe originate from the core-collapse of massive stars ($M> 10 M_{\odot}$), whereas it is believed that type Ia SNe arise
from mass-accretion in binary C-O white dwarfs systems, exploding when the white dwarf mass reaches the Chandrasekhar limit. Two
main scenarios can lead to this situation: i) the single degenerate, where a white dwarf explodes after accreting material from a 
red giant or a main sequence companion (Whelan \& Iben 1973); ii) the double degenerate, where the explosion follows the merger of two C-O white dwarfs (Iben \& Tutukov 1984),
due to loss of angular momentum from gravitational wave emission.
There is not yet consensus on which scenario is better or if both are at work in galaxies. 
In both scenarios type Ia SNe start exploding after 30--40 Myr and continue to do so for a Hubble time. The maximum type Ia SN 
rate depends on the assumed progenitor model but also on the assumed star formation history, being short in systems with a strong burts of star formation (e.g. 0.5 Gyr in the galactic bulge), 1.0-1.5 Gyr in the solar vicinity and several Gyrs in slow star forming systems, such as irregular galaxies (Matteucci \& Recchi, 2001). Therefore, the bulk of Fe is always produced with a delay relative to the $\alpha$-elements produced by type II SNe. 
Manganese is an iron-peak element produced both in type II and Ia SNe.  Nucleosynthesis calculations for massive stars
by Arnett (1971) and Woosley \& Weaver (1995; hereafter WW95), among others, show increased yields of Mn with increasing metallicity
(actually with increasing neutron excess).
Metallicity-dependent yields for type Ia SN have been largely unexplored; however, Ohkubo et al. (2006) 
showed that the yields of both Ni and Mn increase with metallicity in type Ia SNe, although this conclusion is sensitive to the details 
of the adopted hydrodynamical models.

In this paper, we adopt three detailed chemical evolution models for the Galactic bulge, solar vicinity and Sgr dSph.
In all the models we will adopt the same yields but different star formation histories tuned to reproduce the majority 
of the observational constraints on the three systems.
We will show that to explain the [Mn/Fe] simultaneously in the three systems, the
time-delay model alone is not enough, but we have to assume that the yields of Mn from
type Ia SNe are metallicity-dependent.
The paper is organized as follows: in section 2 we briefly describe the observational data, in section 3 we describe the 
chemical evolution models. In section 4 the nucleosynthesis prescriptions are given. In section 5 we discuss our results and in section 6 we draw our conclusions.

\section{Observational data}\label{data}

The observational data used in this paper 
for the Bulge and for the Sgr dSph are taken from 
McWilliam et al (2003). For Sgr dSph a part of
the data come from the work of Bonifacio et al. (2000).
Concerning the Bulge we added the mean values from three
bulge globular clusters measured by Alves-Brito et al. (2005),
Alves-Brito et al. (2006) and Ramirez \& Cohen (2002).
However, it should be noted that to compare the chemical abundances of globular clusters to the results of 
the chemical evolution model of the Bulge is not entirely safe; in fact,
they might have had a different evolution when compared to the evolution 
of the Bulge or the Halo; moreover, it is still uncertain whether these globular clusters 
really belong to the Bulge system.

The data for the solar neighbourhood shown in this paper come from the abundances
reported by Reddy et al. (2003) and Reddy et al. (2006); they are not the only 
data available but we selected them in order to have a very homogeneous sample.

\section{The chemical evolution models}
For studying the chemical evolution of the solar neighbourhood we adopted the model of Fran\c cois et al. (2004),
which is an implemented version of the original model by Chiappini et al. (1997). This model
assumes that the Galaxy formed by means of two main accretion episodes, one giving rise to the halo
and thick disk and the other forming the thin disk.
The infalling gas is always assumed to be of primordial composition. Detailed nucleosynthesis from low 
and intermediate mass stars, type Ia and type II SNe is taken into account. The IMF is taken
from Scalo (1986).  For details of this model see Fran\c cois et al. (2004).

For the Bulge we adopted the model of Ballero et al. (2007), which assumed a rapid formation 
timescale, of 0.3--0.5 Gyr, from gas accumulated during the Halo collapse. The efficiency of star formation (star 
formation per unit mass of gas) is 20 times higher (i.e. $20 Gyr^{-1}$) than in the solar vicinity ($1 Gyr^{-1}$). 
The IMF is flatter than in the solar vicinity, as required by the observed Bulge stellar metallicity distribution. 
See Ballero et al. (2007) for details of this model.

To study Sgr dSph we have adopted the model developed for this galaxy by Lanfranchi et al. (2004; 2006).
This model assumes that Sgr dSph formed stars over several Gyrs, as suggested by its color-magnitude diagram,
and that a strong galactic wind, triggered by supernovae, was responsible for the gas loss.  While the star formation 
efficiency is similar to the solar vicinity, the effect of significant gas loss is similar to
a lower star formation efficiency.
In particular, we assume that the star formation rate is given by a simple Schmidt law:
\begin{equation}
\psi(t)= \nu \sigma_{gas}
\end{equation}
where $\psi(t)$ is the star formation rate, $\nu=1 Gyr^{-1}$ is the star formation efficiency, namely the 
star formation rate per unit mass of gas, and $\sigma_{gas}$ is the surface gas density.
The wind rate is then assumed to be:
\begin{equation}
W(t)=\lambda \psi(t)
\end{equation}
with $\lambda$ being a parameter whose best value for Sgr dSph is $\lambda=13.0$.
The assumed IMF is also flatter than in the solar vicinity (Salpeter 1955).

According to the time delay model, the predicted [$\alpha$/Fe] ratios behave differently in the three systems 
because of the different star formation histories (see Matteucci 2003), in the sense that we expect a long 
plateau with supersolar [$\alpha$/Fe] ratios in the Bulge because of its fast star formation. In fact, in this 
case a lot of Fe is produced by type II SNe before the bulk of it is restored by SNe Ia. The opposite occurs 
for slow star formation systems, such as dwarf spheroidals. In this case the supersolar [$\alpha$/Fe] plateau 
is much shorter than in the solar vicinity since when the type Ia SNe restore the bulk of Fe, the [Fe/H] in the 
interstellar medium is still quite low. This produces a situation where we expect low [$\alpha$/Fe] ratios at 
low metallicities as opposed to the high [$\alpha$/Fe] ratios at high metallicities expected for the Bulge.
The observed behaviour of the [$\alpha$/Fe] ratios in these three systems is in good agreement with these
predictions (e.g. Fran\c cois et al. 2004; Ballero et al. 2007; Lanfranchi et al. 2006).

In the following sections we will test whether a time-delay scenario, where type Ia SNe 
produce the bulk of the Mn (as suggested Gratton 1989) can predict the observed [Mn/Fe]
ratios in these systems.

\section{Nucleosynthesis Prescriptions}{\label{NP}}

As already mentioned, Fe and Mn are produced by type Ia and II SNe.
In our chemical evolution calculations we explore the consequences of three
nucleosynthesis prescriptions: 1. fixed, metallicity-independent, Mn and Fe yields for
type Ia and II SNe, 2. metallicity-dependent type II SNe yields with fixed,
metal-independent, yields for type Ia SNe, and 3. metal-dependent Mn yields 
for both type Ia and type II SNe.



In our first model we utilise the metal-independent yields from Fran\c cois et al. (2004).  
For type II SNe they found that the WW95 results provide the best fit to solar 
vicinity data; in fact, no modifications were required for the yields of Fe 
computed for solar chemical composition.  However, in order to fit the observed 
[Mn/Fe] trend Fran\c cois et al. modified the WW95 Mn yields for type II SNe, 
with a $\sim$70\% increase for 13--18 M$_{\odot}$ progenitors and a $\sim$60\% decrease
for 30--40 M$_{\odot}$ progenitors.  For type Ia SNe Fran\c cois et al. (2004) employed 
the, unmodified, metallicity-independent Iwamoto et al. (1999) yields.  Effectively,
the Fran\c cois et al. (2004) used the age-metallicity relation and modified the Mn yields
as a function of supernova progenitor mass to reproduce the observed trend of [Mn/Fe] with
metallicity in the solar vicinity.  Thus, these yields are fine-tuned to the formation
timescale of the thin disk.  We employed these yields for our initial chemical evolution 
calculations to predict the [Mn/Fe] trend for the metal-independent time-delay scenario.

For our second nucleosynthesis prescription we chose to use the metallicity-dependent 
Mn and Fe yields of WW95 for the type II SNe, and we
adopted the metallicity-independent theoretical yields by Iwamoto et al. (1999)
for the yields from type Ia SNe.

Finally, for the third nucleosynthesis recipe we again employed the WW95 metal-dependent yields
for type II SNe but we also used metal-dependent Mn yields for type Ia SNe.  For the 
metal-dependent type Ia SN Mn yields we modified the values of Iwamoto et al. (1999) by
introducing the following metallicity dependence:

\begin{equation}
Y_{Mn}(z)=Y_{Mn}^{Iwamoto} \left( \frac{z}{z_{\odot}} \right) ^{0.65}
\end{equation}
where $Z$ is the global metal content of the type Ia SN systems at birth.  The form of the 
metal-dependent yields was obtained by requiring that the model predictions fit the observational
data.  Very recently Badenes, Bravo \& Hughes (2008)
found that the Mn/Cr ratio  in the ejecta of type Ia SNe depends on Z, and their modelled 
metallicity-dependent Mn yield is similar 
to that in Equation (1), thus supporting our choice.  Moreover, nucleosynthesis calculations by 
Ohkubo et al. (2006) suggested a dependence on metallicity of the 
yields of Ni and Mn in type Ia SNe.

\begin{figure}[h!]
\begin{center}
\includegraphics[width=0.99\textwidth]{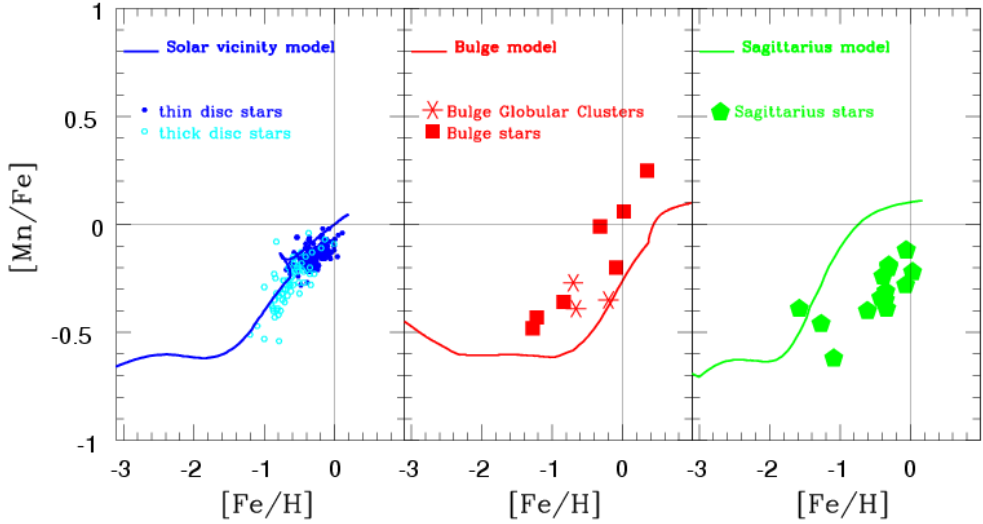}
\caption{The results with Fran\c cois et al (2004) set of yields.
In Fig. the [Mn/Fe] vs [Fe/H] ratios in three different 
stellar systems; from left to right: 
Solar neighbourhood, Bulge and Sagittarius
Dwarf Spheroidal. For the data see  Sect. \ref{data}.
}

\end{center}
\end{figure}

\section{Results}

\subsection{Results with Fran\c cois et al. yields}

In Figure 1 we show the results obtained with the yields of Fran\c cois et al. (2004), and
as expected, the data relative to the stars of the solar vicinity are well reproduced.
Here and in the following figures, the results of all the models
are normalized to the predicted solar abundances,
namely the abundances predicted for the interstellar medium in the solar vicinity 4.5 Gyr ago. 
However, it should be noted that our predicted solar values for most of the considered abundances 
and, in particular for those studied in this paper, differ from the Asplund et al (2005) solar values
by a maximum of $\sim$ 0.06 dex.

On the other hand, Figure 1 shows an unsatisfactory comparison between the time delay model predictions
and the observed Bulge [Mn/Fe] trend with [Fe/H]:
there is an offset between the data and the results of the Bulge 
model, which predicts lower values of the [Mn/Fe] ratio for all metallicities. 

Under the constraint of metallicity-independent Mn yields the time-delay model for
the solar neighbourhood requires that the bulk of the Mn is produced by type Ia SNe.  Because of
the rapid Bulge formation (as suggested by by the alpha-element abundances), the contribution of type Ia SN 
to the bulge composition is smaller than the solar vicinity; therefore, the model predicts less Mn production
in the bulge in this paradigm, and thus lower [Mn/Fe] ratios (contrary to the observations).

Moreover, with this set of yields the results of Sgr dSph model also do not match the observed
[Mn/Fe] trend with [Fe/H]:
the theoretical curve passes through the observed [Mn/Fe] values only at [Fe/H]$\sim-$1.5 dex,
but for higher metallicities the predicted trend is too high compared to the data.  This is
understood as a result of the high frequency of type Ia SNe in the the Sgr dSph model (suggested by 
the low alpha/Fe ratios), which produce more Mn for metallicity-independent Mn yields in the
time-delay scenario. Thus, the prediction of this paradigm is for enhanced [Mn/Fe] ratios in the 
Sgr dSph (again contrary to the observations).

\begin{figure}[h]
\begin{center}
\includegraphics[width=0.99\textwidth]{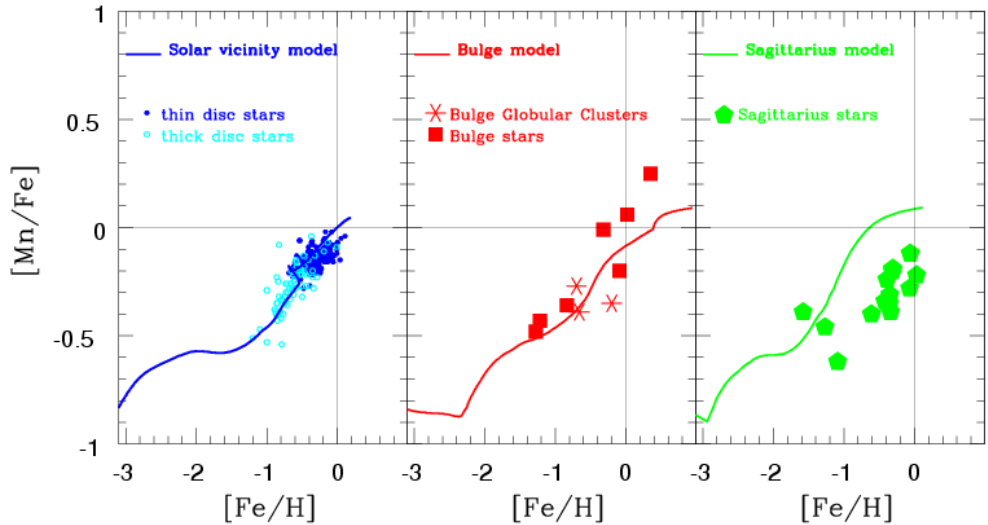}
\caption{The results from the WW95 metallicity-dependent
type II SN yields, with constant, Iwamoto et al. (1999) type Ia SN Mn yields.
The [Mn/Fe] versus [Fe/H] ratios in three different 
stellar systems, from left to right: 
Solar neighbourhood, Bulge and Sagittarius
Dwarf Spheroidal galaxy. For the data see  Sect. \ref{data}.}
\end{center}
\end{figure}

\subsection{Results with Woosley \& Weaver yields}

To improve the match between our models and the data, in particular for the Bulge
and Sgr dSph, we now explore the use of the metallicity-dependent WW95 yields
from type II SNe, but with the constant (i.e. metallicity-independent) Mn and Fe
yields for type Ia SNe from Iwamoto et al. (1999); the results are shown in Figure 2.

The results of the solar neighbourhood model still follow nicely
the observational data.

The model results for the Bulge are also more promising; in fact, up to solar [Fe/H]
the predictions well trace the observational data. However, for supersolar
metallicities the results of the model tend to slightly underestimate the value 
on [Mn/Fe] in the Bulge stars.  Because of the high star formation rate in the Bulge
type Ia SNe occur at late times (and high [Fe/H]). Thus, the Mn deficit
at supersolar metallicities in this Bulge model might reasonably be due to a
problem with Mn from type Ia SNe.

The results from this model for the Sgr dSph are still in poor agreement with the
data. The model shows a strong increase of the [Mn/Fe] ratio at [Fe/H]$\sim-$1.5
dex, at variance with the data.  Given the prominent role of type Ia SNe in Sgr dSph
this Mn over-prediction is likely related to the type Ia SN Mn yields.

\begin{figure}[h]
\begin{center}
\includegraphics[width=.99\textwidth]{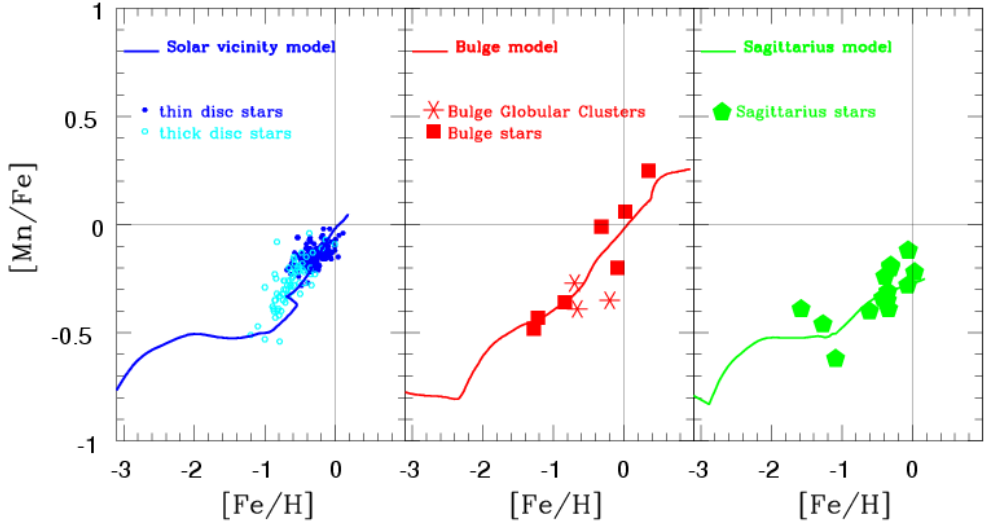}
\caption{The results with metallicity dependence in both type II SNe and type Ia SN 
yields for Mn.
In Fig. the [Mn/Fe] vs [Fe/H] ratios in three different 
stellar systems; from left to right: 
Solar neighbourhood, Bulge and Sagittarius
Dwarf Spheroidal. For the data see  Sect. \ref{data}.}
\end{center}
\end{figure}

\subsection{Results with Woosley \& Weaver yields and metallicity dependent Mn yields from type Ia SNe}
In this last set of chemical evolution models both the type Ia and II SN yields of Mn are
dependent on the metallicity. In particular, the Mn yields from type Ia SNe follow 
Equation (1).  In this case the yields are constrained to maintain agreement between the chemical 
evolution model predictions for [Mn/Fe] versus [Fe/H] and the solar vicinity observational data.
At the same time, the supersolar data in the Bulge are nicely fit.
Thus, the use of this new set of yields had improved the predictions of
the Bulge model, presumably because the Mn contribution from late time, high metallicity, 
type Ia SNe are more correctly modelled.
Finally, the most interesting result is the excellent agreement
of the Sgr dSph model with the data. This is due to the reduced
contributions of the Mn enrichment by low metallicity type Ia SNe, compared to the
time-delay metallicity-independent model.

A very important conclusion is that the yield of Mn produced
by type Ia SN {\em must} be metallicity-dependent in order to fit
the Sgr dSph abundance data. Moreover, we get excellent agreement also for the Bulge and 
solar vicinity stars. 
It is worth emphasising that it was not possible to obtain this result 
without the simultaneous comparison of the three independent stellar systems, charcaterised
by different star formation timescales.
This hypothesis, that the type Ia SN Mn yield {\em must} be dependent
on metallicity, was first suggested by McWilliam et al. (2003). 

\section{Conclusions}
We show quantitatively, for the first time, that galactic chemical evolution models demand 
that the type Ia SN Mn yields must be metallicity-dependent, by
simultaneously fitting the [Mn/Fe] trend in three 
independent stellar systems: the Galactic bulge, 
the solar vicinity and the Sgr dSph. 
We have adopted detailed chemical evolution models, chosen to reproduce the main
chemical features of these three stellar systems, using same element yields but different 
star formation histories.

We have shown clearly that the ``time-delay model'' alone cannot explain the different 
behaviour of the [Mn/Fe] in Bulge, solar vicinity and Sgr dSph,
as it does for [$\alpha$/Fe] ratios. In fact, in the case of the [$\alpha$/Fe] ratios,
the same models can well reproduce the observed [$\alpha$/Fe] versus [Fe/H] trends in the
three stellar systems without invoking metallicity-dependent yields. This is because 
the production of $\alpha$ and Fe elements is only mildly dependent on the initial stellar 
metallicity.  We propose a functional form of the metallicity-dependent Mn yields in
type Ia SNe ($ \propto Z^{0.65}$) which should be tested by nucleosynthesis calculations.
Our result is supported by Ohkubo et al. (2006), who suggested that the yields of Mn from 
type Ia SNe have a strong dependence on metallicity, and by a recent finding by Badenes et al. (2008),
who showed that the Mn to Cr ratio in the ejecta of type Ia SNe depend on the metallicity
of the progenitor star in a very similar way to what we have suggested.

It is worth noting that in this work we have taken the Reddy et al. (2003, 2006) abundances for the
solar vicinity stars at
face value.  However, we note that McWilliam et al. (2003) found that a
zero-point correction of $+$0.10 dex to the Reddy et al. (2003) [Mn/Fe] values
was required if the sun is typical of stars in the range 
$-$0.10 $\leq$ [Fe/H] $\leq$< $+$0.10.  We have re-investigated this issue
and found that together the Reddy et al. (2003) and Reddy et al. (2006) data
indicate a mean [Mn/Fe] of solar [Fe/H] stars at $-$0.10$\pm$0.01 dex, with
an rms scatter about the mean of 0.036 dex.  Since the solar Mn/Fe ratio is 
2.77 sigma above the mean for solar [Fe/H] stars (equal to 99.7\% above the mean
in a gaussian distribution), this zero-point shift is most likely due to systematic 
error in the Reddy et al. (2003, 2006) results, rather than a real Mn enhancement in
the Sun.  This is supported by the fact that other abundance studies require quite 
different zero-point corrections; for example the differential study, relative to the
sun, by Feltzing \& Gustafsson (1998) required a zero-point [Mn/Fe] correction of 
$-$0.06 dex.

Unfortunately, we are unable to make as good a fit to the measured [Mn/Fe]
trends if we employ a 0.10 dex zero-point shift to the Reddy et al. (2003, 2006)
data.  To do so would require a more complicated form of the metallicity-dependent 
Mn yields from type Ia SNe, instead of the simple power-law form adopted here.
Alternatively, other complications might also be responsible, such as the type II
SNe yields of WW95, or variations in the star formation and/or the mass-loss rates
for Sgr dSph.  Rather than explore a myriad of possibilities that might give a perfect
fit to all the Mn trends, we prefer to focus on our main quantitative result: that
the trend of [Mn/Fe] with [Fe/H] in the Galactic disk, the Bulge, and the Sgr dSph
can only be understood if the yield of Mn from type Ia SNe is metallicity-dependent.

Finally, we note that the low [Mn/Fe] values seen in Sgr dSph stars near solar metallicity is
due, at least in part, to the lower metallicity of old Sgr dSph stars compared to the
solar neighborhood.  It is the low average [Fe/H] of
the type Ia SNe in the Sgr dSph that results in lower [Mn/Fe] values, thanks to the
metallicity-dependence of the Mn yield.  Thus, the [Mn/Fe] ratios of solar metallicity
stars in dwarf galaxies is linked to the mean metallicity of these galaxies. Since the
metallicity distribution function of stars in dwarf galaxies is sensitive to the mass
outflow and star formation rates, the value of these parameters should be constrained
by the observed [Mn/Fe] ratios.  Indeed, we would expect that the more massive dwarf
galaxies (e.g. LMC) would show a similar effect on [Mn/Fe], but less extreme, than in lower
mass dwarf galaxies; thus, there should be families of low-mass galaxies with different 
[Mn/Fe] trends.


\section{Acknowledgments}
G.C. acknowledges financial support from the Fondazione Cassa di Risparmio di Trieste.


\begin{thebibliography}{1000}

\bibitem {alves-b1}
Alves-Brito, A., Barbuy, B., Ortolani, S., et al.
2005 A\&A, 435, 657,

\bibitem {alves-b2}
Alves-Brito, A., Barbuy, B., Zoccali, M., et al. 
2006, A\&A, 460, 269

\bibitem{Arnett}
Arnett, W.D. 1971, ApJ, 169, 113

\bibitem[\protect\citeauthoryear{Asplund et al.}{2005}]{b41}
Asplund, M., Grevesse, N., Sauval, A.J.
2005, ASPC, 336, 25A	


\bibitem {Badenes}
Badenes, C., Bravo, E., Hughes, J.P.
2008, ApJ, 680L, 33

\bibitem[\protect\citeauthoryear{Ballero et al.}{2007}]{b10}
Ballero, S.K., Matteucci, F., Origlia, L., Rich, R.M. 
2007, A\&A, 467, 123

\bibitem[\protect\citeauthoryear{Bonifacio et al.}{2000}]{b20}
Bonifacio, P., Hill, V., Molaro, P., et al.
2000, A\&A, 359, 663

\bibitem[\protect\citeauthoryear{Chiappini et al.}{1997}]{110} 
  Chiappini, C., Matteucci, F., Gratton, R.G.,
  1997, ApJ, 477, 765

\bibitem[\protect\citeauthoryear{Feltzing et al.}{1998}]{f2}
Feltzing, S. \& Gustafsson, B. 1998, A\&AS, 129, 237

\bibitem[\protect\citeauthoryear{Fran\c cois et al.}{2004}]{b30}
Fran\c cois, P., Matteucci, F., Cayrel, R., et al.
2004, A\&A, 421, 613

\bibitem{Gratton}
Gratton, R.G. 1989, A\&A 208, 171

\bibitem[\protect\citeauthoryear{Greggio \& Renzini}{1983}]{b23}
Greggio, L., Renzini, A. 1983, MmSAI, 54, 311 

\bibitem{Iben1}
Iben, I.Jr. \& Tutukov, A.V. 1984, ApJS, 54,335

\bibitem{Iwamoto}
Iwamoto, K., Brachwitz, F., Nomoto, K., et al.
1999, ApJS, 125, 439 
 
\bibitem{Lanfranchi2}
Lanfranchi, G.A., Matteucci, F. 2004, MNRAS, 351, 1338

\bibitem{Lanfranchi3}
Lanfranchi, G.A., Matteucci, F., Cescutti, G. 2006, MNRAS, 365, 477


\bibitem[\protect\citeauthoryear{Matteucci \& Greggio}{1986}]{b320} 
Matteucci, F., Greggio, L. 1986, A\&A, 154, 279

\bibitem[\protect\citeauthoryear{Matteucci \& Recchi}{2001}]{b24} 
Matteucci, F., Recchi, S. 2001, ApJ, 558, 351

\bibitem[\protect\citeauthoryear{Matteucci }{2003}]{b25} 
Matteucci, F. 2003, Ap\&SS, 284, 539


\bibitem[\protect\citeauthoryear{McWilliam et al.}{2003}]{b50}
McWilliam, A., Rich, R.M., Smecker-Hane, T.A.
2003, ApJ, 592L, 21

\bibitem{ohkubo}
Ohkubo, T., Umeda, H., Nomoto, K., Yoshida, T. 2006, AIPC, 847, 458

\bibitem{ramirez}
Ram\'irez, S.V., Cohen, J.G. 
2002, AJ, 123, 3277

\bibitem[\protect\citeauthoryear{Reddy et al.}{2003}]{b60}
Reddy, B.E., Tomkin, J., Lambert, D.L., Allende Prieto, C.
2003, MNRAS, 340, 304

\bibitem[\protect\citeauthoryear{Reddy et al.}{2003}]{b70}
Reddy, B.E., Tomkin, J., Lambert, D.L., Allende Prieto, C.
2006, MNRAS, 367, 1329

\bibitem[\protect\citeauthoryear{Salpeter}{1955}]{S55}
Salpeter E.E. 1955, ApJ, 121, 161

\bibitem{tinsley}
Tinsley, B.M.
1979, ApJ, 229, 1046

\bibitem[\protect\citeauthoryear{Whelan}{1973}]{WI}
Whelan, J. \& Iben, I.Jr. 1973, ApJ, 186, 1007

\bibitem[\protect\citeauthoryear{Woosley \& Weaver.}{1995}]{b80} 
Woosley, S.E., Weaver, T.A.
1995, ApJ, 101, 181 


\end{thebibliography}
\end{document}